\documentclass[amsmath,superscriptaddress,showpacs,prx,twocolumn,floatfix]{revtex4-1}
\usepackage{graphicx}
\usepackage{multirow}
\usepackage{latexsym,amssymb}
\usepackage{amsmath,bbold}
\usepackage{paralist}
\usepackage{braket}

\newcommand{\del}[2]{\partial_{#2}{ #1}}

\newcommand{\dprime}[1]{#1^{\prime\prime}}
\newcommand{\BDG}[1]{Bogoliubov-deGennes }
\DeclareMathAlphabet{\mathpzc}{OT1}{pzc}{m}{it}

\begin{document}

\title{Regular and in-plane skyrmions and antiskyrmions from boundary instabilities}
\author{Shane Sandhoefner}
\affiliation{Department of Physics and Astronomy and Nebraska Center for Materials and Nanoscience, University of Nebraska, Lincoln, Nebraska 68588, USA}

\author{Aldo Raeliarijaona}
\affiliation{Department of Physics and Astronomy and Nebraska Center for Materials and Nanoscience, University of Nebraska, Lincoln, Nebraska 68588, USA}

\author{Rabindra Nepal}
\affiliation{Department of Physics and Astronomy and Nebraska Center for Materials and Nanoscience, University of Nebraska, Lincoln, Nebraska 68588, USA}

\author{Dalton Snyder-Tinoco}
\affiliation{Department of Physics, California State University at San Bernardino, San Bernardino, California 92407, USA}

\author{Alexey A. Kovalev}
\affiliation{Department of Physics and Astronomy and Nebraska Center for Materials and Nanoscience, University of Nebraska, Lincoln, Nebraska 68588, USA}
\date{\today}

\begin{abstract}
We formulate a theory of skyrmion and antiskyrmion generation using magnetic field and charge current pulses. We show that the topological defect can be created at an edge of a system with Dzyaloshinskii-Moriya interaction (DMI) as well as at a boundary between regions with different DMI. We consider both perpendicular and in-plane (also known as
magnetic bimerons) versions of skyrmions and antiskyrmions. We show that the magnetization twist in the vicinity of an edge or a boundary is described by a kink solution, the presence of which can instigate the generation of topological defects. We study the collective excitations of magnetization analytically and numerically, and demonstrate that under application of magnetic field and charge current pulses the magnon modes localized near boundaries can develop instabilities leading to the formation of skyrmions or antiskyrmions. Due to the skyrmion and antiskyrmion Hall effects, a properly chosen current direction can push the topological defects away from the boundary, thus facilitating their generation.
\end{abstract}

\maketitle

\section{Introduction} Magnetic skyrmions and antiskyrmions, which are topologically protected whirls of magnetic moments on the nanometer scale, have been a topic of great interest in recent years \cite{Bogdanov.YablonskiiJETP1989,Mbauer,Bogdanov.Roesler.eaPRB2002,Robler-Bogdanov-Pfeiderer,Everschor2018,Fert-Review,Nagaosa-Review,Koshibae.Nagaosa:NC2016}. The Dzyaloshinskii-Moriya interaction (DMI) helps stabilize these structures, among other mechanisms such as dipole-dipole interaction \cite{Desautels2019} or the competing exchange interactions between neighbors \cite{Okubo2012,Leonov2015}. The symmetry or asymmetry of interfacial DMI determines the type of structure formed \cite{PhysRevB.93.064428,Hoffmann}. For instance in chiral magnets, Rashba-type DMI leads to skyrmions and Dresselhaus-type DMI leads to antiskyrmions~\cite{10.3389/fphy.2018.00098}. In addition to their fundamental interest, there are proposals to use skyrmions or antiskyrmions in memory devices \cite{Ruff,XZhang} and reservoir computing \cite{Prychynenko2018}. Due to different conditions required to create stable skyrmions or antiskyrmions, methods for their generation form an important piece of skyrmion related research. There are a number of theoretical proposals \cite{PhysRevApplied.11.024051,Deger2019,PhysRevMaterials.2.124401,Muller_2016,Yuan2016}, as well as direct experimental observations of skyrmion generation~\cite{Desautels2019,Tomasello2014,Woo2016,Moreau-Luchaire2016,Boulle2016,Soumyanarayanan2017}. It is desirable to develop universal means for generating both skyrmions and antiskyrmions. 

Layered magnetic heterostructures suitable for realizations of skyrmions or antiskyrmions typically contain perpendicular magnetocrystalline anisotropy and have a perpendicular magnetization configuration. In-plane skyrmions or antiskyrmions (also known as magnetic bimerons)~\cite{PhysRevLett.119.207201}, on the other hand, can be realized in systems with in-plane magnetization and in-plane anisotropy~\cite{PhysRevB.99.060407,PhysRevLett.124.037202,PhysRevB.101.054405}.
Realizations of in-plane skyrmions also require a special form of DMI component proposed in Ref.~\cite{PhysRevB.93.064428} for systems with only mirror symmetry, which can stabilize spirals with preferred in-plane configuration~\cite{PhysRevB.93.064428}. As shown in Refs.~\cite{PhysRevB.99.060407,PhysRevLett.124.037202,PhysRevB.101.054405,PhysRevB.101.054405}, such DMI can also lead to realizations of in-plane skyrmions or antiskyrmions in monoclinic systems described by the point group $Cm$ with only mirror symmetry. 

The edges or boundaries can become preferable locations for generation of skyrmions or antiskyrmions as DMI causes magnetization near edges or boundaries to twist \cite{Wilson.Karhu.eaPRB2013,Rohart-Thiaville,Du.Che.eaNC2015,Meynell.Wilson.ea:PRB2014}.  In the presence of edge or boundary instabilities, chiral domain walls can form from the twist of magnetization at the edge or boundary and evolve into skyrmions or antiskyrmions. Previous studies have investigated the possibility of skyrmion or antiskyrmion generation at edges or boundaries through application of magnetic field pulse~\cite{Muller_2016,PhysRevMaterials.2.124401}. Generation of skyrmions in the bulk by charge current pulses has also been proposed~\cite{Stier}. 

In this paper, we expand upon the above ideas by considering the charge current. We formulate a theory of regular and in-plane skyrmion and antiskyrmion generation at edges of magnetic films and at boundaries between regions with different DMI. The process of generation is triggered by local instabilities at edges or boundaries due to lowering of magentic field or application of charge current pulse. To identify the appearance of instabilities, we study
the magnon modes localized at edges or boundaries. 
By studying a charge current flowing along the edge or boundary, we observe that the presence
of the skyrmion or antiskyrmion Hall effect~\cite{Iwasaki2013,Woo2016} can facilitate
the generation of topological defects. Depending on the direction of charge current, topological defects are pushed either away from or towards the edge or boundary, which either facilitates or suppresses the generation of skyrmions or antiskyrmions.

The paper is organized as follows. In Section II, we discuss the boundary conditions for chiral ferromagnets and describe the magnetization twists that can arise at edges of magnetic films or at boundaries between regions with different DMI. We also formulate the Bogoliubov-de Gennes Hamiltonian describing magnon modes localized on magnetization twists. In Section III, we study the magnon gap for modes localized on edges or boundaries and identify instabilities associated with the closure of the magnon gap. Using micromagnetic simulations, we show that such instabilities can lead to generation of skyrmions and antiskyrmions. We summarize our results in Section IV.

\section{Method} \subsection{Free energy and boundary conditions} We consider a chiral ferromagnet well below the Curie temperature with a free energy density:
\begin{equation}\label{FreeE}
{\mathcal F}=(J/2) (\partial_j\boldsymbol m)^2+ \boldsymbol{\mathcal{D}}_j \cdot (\partial_j\boldsymbol m \times  \boldsymbol m) -m^T \hat{K} m-\boldsymbol H m,
\end{equation}
where $\boldsymbol m$ describes a unit vector along the magnetization direction and summation over repeated indices is assumed. The term $J$ describes exchange stiffness, the term $\hat{K}$ describes magnetic anisotropy, and the term $\boldsymbol{\mathcal{D}}_j$ describes DMI, $(\boldsymbol{\mathcal{D}}_j)_i=\mathcal{D}_{ij}$, where $\mathcal{D}_{ij}$ is the rank-2 DMI tensor. The magnetic field term $\boldsymbol H$ includes both the external, $\boldsymbol H_e$, and the dipolar, $\boldsymbol H_d$, magnetic fields, $\boldsymbol H\equiv\mu_{0}M (\boldsymbol H_e + \boldsymbol H_d)$. For a thin magnetic film, the dipolar magnetic fields due to normal to the film magnetization can be included into the effective shape anisotropy $\hat{K}_{eff}$~\cite{PhysRev.124.452}.  In our analytical results, we do not consider dipolar interactions, but micromagnetic calculations are performed both in the absence and in the presence of dipolar interactions.

We consider a system with a boundary between regions of differing DMI and assume that the directions of the magnetic field and magnetic anisotropy do not change across the boundary. Using the variational principle, one can obtain the boundary conditions \cite{PhysRevMaterials.2.124401},
\begin{equation}\label{BC2}
 J^{(1)} n^{(1)}_j \del{\boldsymbol m}{j}+J^{(2)} n^{(2)}_j \del{\boldsymbol m}{j}+\boldsymbol \Gamma^{(1)}_D+\boldsymbol \Gamma^{(2)}_D =0,
\end{equation}
where $\boldsymbol n^{(1)}$, $J^{(1)}$, and $\boldsymbol \Gamma^{(1)}_D$ correspond to the first region and $\boldsymbol n^{(2)}$, $J^{(2)}$, and $\boldsymbol \Gamma^{(2)}_D$ correspond to the second region. Here $\boldsymbol n^{(i)}$ is the normal pointing outside of the region, and $(\Gamma^{(i)}_{D})_k=m_i n^{(i)}_j (\epsilon^{kmi}\mathcal{D}^{(i)}_{mj})$ with $\epsilon^{kmi}$ being the Levi-Civita symbol. For an edge with vacuum, this reduces to \cite{Hals.Everschor-Sitte:PRL2017}
\begin{equation}\label{BC3}
 J n_j \del{\boldsymbol m}{j}+\boldsymbol \Gamma_D =0,
\end{equation}
with $(\Gamma_{D})_k=m_i n_j (\epsilon^{kmi}\mathcal{D}_{mj})$.

\begin{figure}
\centering
\includegraphics[width=\linewidth]{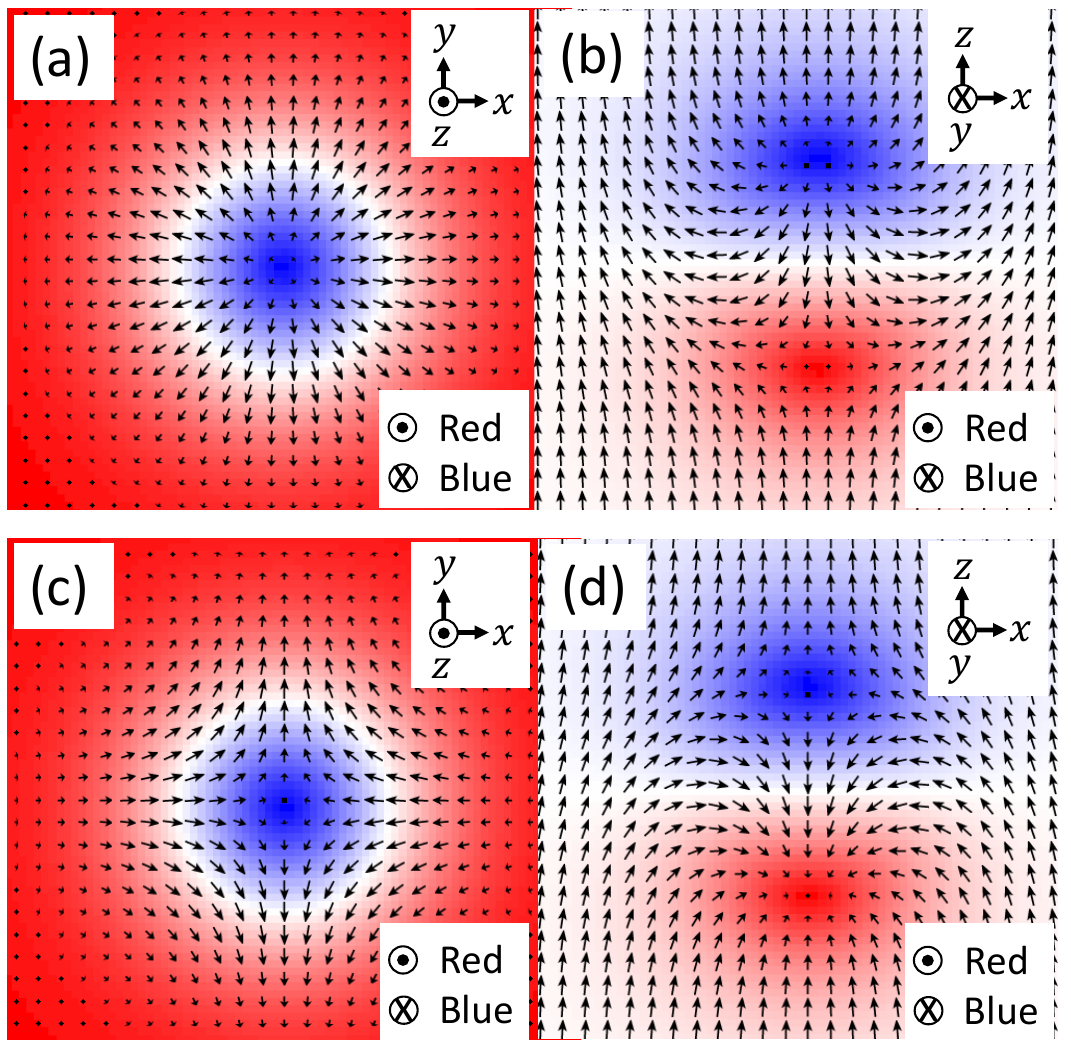}
\caption{(Color online) (a) Regular skyrmion, (b) in-plane skyrmion, (c) regular antiskyrmion, and (d) in-plane antiskyrmion, where the in-plane (anti)skyrmion is obtained from the regular (anti)skyrmion by sending $x$ to $x$, $y$ to $z$, and $z$ to $-y$.}
\label{fig:IntroBimeron}
\end{figure}

Below, we consider a quasi-two-dimensional ferromagnet. We analyze forms of the free energy related to each other by a global transformation in the spin space applied to the magnetization~\cite{PhysRevB.93.064428}. As long as such transformations are applied to all vectors and tensors entering the free energy density, the value of the free energy density (and all related physics) is preserved~\cite{PhysRevB.93.064428}. We write the same free energy density to describe regular and in-plane skyrmions and antiskyrmions:
\begin{equation}\label{FreeE2}
{\mathcal F}=\frac{J}{2}(\partial_{\alpha}\boldsymbol m)^2 -K m_{z}^{2}-H m_{z}+ \boldsymbol{\mathcal{D}}_j\cdot(\partial_j\boldsymbol m \times  \boldsymbol m).
\end{equation}
In particular, to describe skyrmions and antiskyrmions, we assume that a quasi-two-dimensional ferromagnet is in the $x-y$ plane. To describe in-plane skyrmions and antiskyrmions, we assume that a quasi-two-dimensional ferromagnet is in the $x-z$ plane, see Fig.~\ref{fig:IntroBimeron}. This effectively corresponds to a global transformation in the spin space with rotation by $90^{\circ}$ around the $x$-axis.

To uncover relevant to our discussion physics, we consider the following DMI parametrization~\cite{PhysRevMaterials.2.124401}:
\begin{eqnarray}\label{DMIGeneral}
\boldsymbol{\mathcal{D}}_{1}=\{D_1,D_2,0\},\\
\boldsymbol{\mathcal{D}}_{2}=\{D_3,D_4,0\},\\
\boldsymbol{\mathcal{D}}_{3}=\{D_3,D_4,0\},
\end{eqnarray} 
where $\boldsymbol{\mathcal{D}}_{1}$ and $\boldsymbol{\mathcal{D}}_{2}$ are used to describe a quasi-two-dimensional ferromagnet in the $x-y$ plane and  $\boldsymbol{\mathcal{D}}_{1}$ and $\boldsymbol{\mathcal{D}}_{3}$ are used to describe a quasi-two-dimensional ferromagnet in the $x-z$ plane. Our results obtained for skyrmions or antiskyrmions will also apply to their in-plane versions in magnetic systems with in-plane magnetization, and vice versa. We note in passing that to describe the Rashba-type DMI we choose $D_1=D_4=0$ and $D_2=-D_3$, and to describe the Dresselhaus-type DMI we choose $D_1=D_4=0$ and $D_2=D_3$.

\subsection{Boundary magnetization twists}
\begin{figure}
\centering
\includegraphics[width=\linewidth]{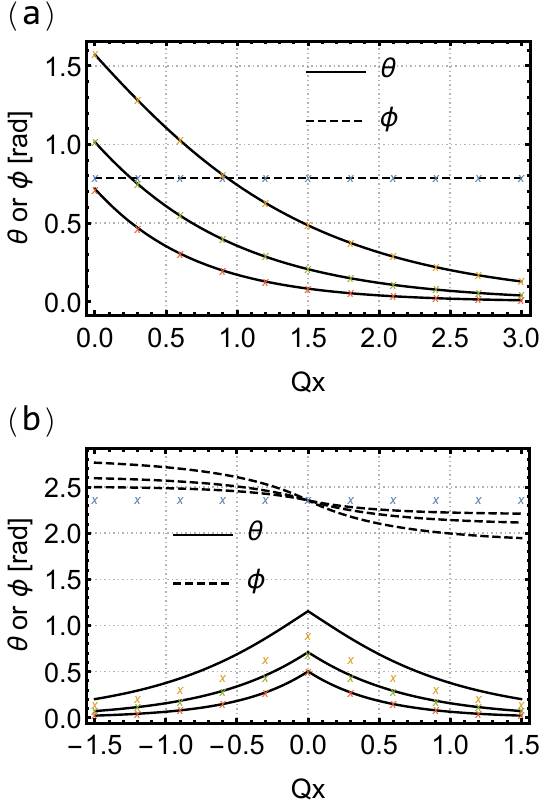}
\caption{(Color online) Lines represent magnetization profiles in the vicinity of (a) an edge and (b) a boundary, obtained by micromagnetic simulations for parameters $J=30~$pJ$/$m and $K=9\times10^4\,$J$/$m$^3$. The corresponding analytical results from Eqs.~\eqref{DSG2}-\eqref{eq:phi2} are shown by crosses. The bold and dashed lines correspond to the magnetic fields $H_e=0.8$~T, $0.3$~T, and $0.1$~T for the lower, middle, and upper curves, respectively. In (a), we use $D_1=-D_2=2.1~$mJ$/$m$^2$ corresponding to the solution $\phi=\pi/4$ shown by the dashed line. In (b), we use $D^{(1)}_1=0$, $D^{(1)}_2=-3$~mJ$/$m$^2$, $D^{(2)}_1=3$~mJ$/$m$^2$ $D^{(2)}_2=0$. The dashed lines represent variation of $\phi$, with the largest variation corresponding to the smallest magnetic field strength.}
\label{fig:twists}
\end{figure}

The equilibrium magnetization profile in the vicinity of a boundary can be obtained by minimizing the free energy. Without loss of generality, we consider a boundary normal to the $x$-axis at $x=0$. We use spherical coordinates, i.e., $\boldsymbol m=[\sin\theta \cos\phi,\sin\theta \sin\phi,\cos\theta]$, where $\theta$ is the polar angle and $\phi$ is the azimuthal angle with respect to the $z$-axis. This results in the following Euler-Lagrange equations: 
\begin{align}\label{DSG}
J\dprime{\theta}-K\sin[2\theta(x)]&-H \sin[\theta(x)]\\ \nonumber
&=\frac{J}{2}\sin[2\theta(x)]\phi^{\prime^2}-2\Tilde{D}\sin[\theta(x)]^2\phi^\prime,\\
J\sin[\theta(x)]\dprime{\phi}+2J &\cos[\theta(x)] \phi^\prime \theta^\prime-2\Tilde{D}\sin[\theta(x)]\theta^\prime=0,\label{DSG1}
\end{align}
where $\Tilde{D}=D_1 \cos[\phi(x)]+D_2\sin[\phi(x)]$. Equations \eqref{DSG} and \eqref{DSG1} lead to the double Sine-Gordon equation:
\begin{align}\label{DSG2}
J\dprime{\theta}-K\sin[2\theta(x)]&-H \sin[\theta(x)]=0,
\end{align}
under assumptions $\phi=\text{const}$ and $\Tilde{D}=0$. These assumptions do not hold for all shapes of DMI tensor (see, e.g., Fig.~\ref{fig:twists} obtained using Mumax3~\cite{Vansteenkiste2014}). However, the assumptions hold for constrained DMI tensors, $\mathcal{D}_{ij}$, with either $D_1=D_4=0$ or $D_2=D_3=0$, as in these two cases the boundary conditions take the following form:
\begin{align}
J^{(1)}\theta^\prime\rvert_{0-}&-J^{(2)}\theta^\prime\rvert_{0+}=\sqrt{(\Delta D_1)^2+(\Delta D_2)^2},\label{sinphi}\\
\sin\phi=&\frac{\Delta D_1}{\sqrt{(\Delta D_1)^2+(\Delta D_2)^2}},\label{sin1}\,\\
\cos\phi=&-\frac{\Delta D_2}{\sqrt{(\Delta D_1)^2+(\Delta D_2)^2}}\label{sin2},
\end{align}
where $\Delta D_1=D_1\rvert_{0+}-D_1\rvert_{0-}$ and $\Delta D_2=D_2\rvert_{0+}-D_2\rvert_{0-}$ describe the change of DMI across the boundary. For a non-constrained DMI tensor, conditions \eqref{sin1} and \eqref{sin2} only approximately determine the angle $\phi$, as can be seen in Fig.~\ref{fig:twists}(b). For an edge with vacuum at $x<0$, the boundary conditions reduce to
\begin{align}\label{sinphi1}
-J\theta^\prime\rvert_{0+}=&\sqrt{(D_1)^2+(D_2)^2},\\
\sin\phi=&\frac{D_1}{\sqrt{(D_1)^2+(D_2)^2}},\label{eq:phi1}\\
\cos\phi=&-\frac{D_2}{\sqrt{(D_1)^2+(D_2)^2}}.\label{eq:phi2}
\end{align}
Note that for an edge with vacuum, the conditions $\phi=\text{const}$ and $\Tilde{D}=0$ can be satisfied irrespective of the form of the DMI tensor, $\mathcal{D}_{ij}$, as can be seen from Eqs.~\eqref{DSG} and \eqref{DSG1}.  

We introduce the dimensionless units for the length, e.g., $x$ is redefined as $Q^{(i)}x$, with $Q^{(i)}=\sqrt{(\Delta D_1)^2+(\Delta D_2)^2}/2J^{(i)}$. We also introduce dimensionless $h^{(i)}=H^{(i)}/[J^{(i)} (Q^{(i)})^2]$, $\kappa^{(i)}=2K^{(i)}/[J^{(i)} (Q^{(i)})^2]$, and $d^{(i)}_{k}=D^{(i)}_{k}/[J^{(i)} Q^{(i)}]$.
For an edge with vacuum, we introduce $Q=\sqrt{( D_1)^2+(D_2)^2}/J$, $h=H/[J Q^2]$, $\kappa=2K/[J Q^2]$, and $d_{k}=D_{k}/[J Q]$.
The double Sine-Gordon equation \eqref{DSG2} can be solved by multiplying with $\theta^\prime$ and integrating with the boundary conditions $\theta(\pm\infty)=\theta^\prime(\pm\infty)=0$. We thus obtain the kink solution \cite{PhysRevB.27.474,Muller_2016}, where a twist in the magnetization can be described by 
\begin{equation}\label{ThetaInt}
\theta(x)=\pi- 2\tan^{-1}\left({\frac{\sqrt{h} \sinh\left\lbrace\sqrt{h+\kappa}\left(|x|-x_0\right)\right\rbrace}{\sqrt{h+\kappa}}} \right),
\end{equation}
where the boundary is at $x=0$ and $x_0$ is the coordinate of the kink. 
To find $x_0$, one can use the boundary condition, Eq.~\eqref{sinphi}, which in dimensionless units becomes:
\begin{equation}
\theta^\prime\rvert_{0-}-\theta^\prime\rvert_{0+}=2.    
\end{equation}
For the symmetric case, i.e. when $h=h^{(1)}=h^{(2)}$ and $\kappa=\kappa^{(1)}=\kappa^{(2)}$, this leads to
\begin{equation}\label{x0Edge}
x^{(i)}_0=-(-1)^i\frac{\cosh^{-1}{\bigg(\frac{(h+\kappa)+\sqrt{(h+\kappa)^2-\kappa}}{\sqrt{h}}}\bigg)}{\sqrt{(h+\kappa)}}.
\end{equation}
For an edge with vacuum, we use Eq.~\eqref{sinphi1},
\begin{equation}
\theta^\prime\rvert_{0+}=-1,    
\end{equation}
which leads to
\begin{equation}\label{x0Edge1}
x_0=-\frac{\cosh^{-1}{\bigg(\frac{(h+\kappa)+\sqrt{(h+\kappa)^2-\kappa}}{\sqrt{h}}}\bigg)}{\sqrt{(h+\kappa)}}.
\end{equation}

In Fig.~\ref{fig:twists}, we show magnetization profiles close to an edge and a boundary obtained by micromagnetic simulations. The edge profile in Fig.~\ref{fig:twists}(a) is in agreement with the analytical results in Eqs.~\eqref{eq:phi1}-\eqref{ThetaInt}. On the other hand, the boundary profile in Fig.~\ref{fig:twists}(b) exhibits deviations from analytical results in Eqs.~\eqref{DSG2}-\eqref{sin2} due to variations in angle $\phi$. 

\begin{figure}
\centering
\includegraphics[width=\linewidth]{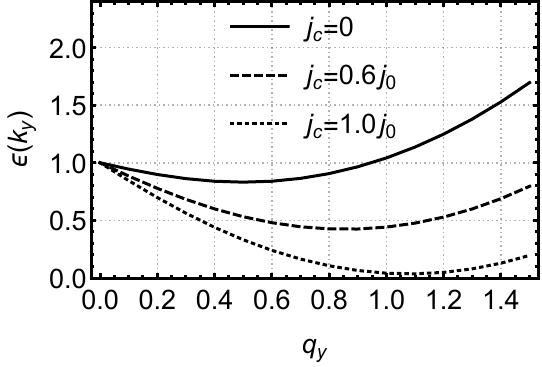}
\caption{The magnon spectrum of the lowest energy mode localized at the edge for different values of the current. We use the dimensionless units with parameters given by $\kappa=0.5$, $h=0.5$, and $j_0=\frac{ 2eJQ}{\hbar{\cal P}}$.}
\label{fig:modes}
\end{figure}

\subsection{Bogoliubov-de Gennes Hamiltonian}
\label{section:LLG equation}
To study the current-induced instabilities, we utilize the Landau-Lifshitz-Gilbert (LLG) equation amended by the current-induced torque term, \cite{PhysRevB.93.064428,PhysRevB.99.060407,PhysRevB.101.054405}
\begin{align}\label{LLG}
s(1-\alpha\boldsymbol m \times) \dot{\boldsymbol m} - \boldsymbol m \times \delta_{\boldsymbol m} F =\boldsymbol \tau,
\end{align}
where $s=M_s / \gamma$ is the spin angular momentum density, $M_s$ is the saturation magnetization, $\gamma$ is 
(minus) the gyromagnetic ratio ($\gamma>0$ for electrons), $\alpha$ is the Gilbert damping, $F$ is the total free energy, and $\boldsymbol \tau=\boldsymbol \tau_{st}+\boldsymbol \tau_{so}$ describes the spin-transfer and spin-orbit torques. The spin-transfer torque is given by $\boldsymbol \tau_{st}=(\hbar {\cal P})/(2e)(1-\beta \boldsymbol m \times) (\boldsymbol j_c \cdot \boldsymbol  \nabla ) \boldsymbol m$, where $e>0$ is (minus) the electron charge, $j_c$ is the charge current density, $\beta$ is the factor describing non-adiabaticity, and $\cal P$ is the efficiency of the spin-transfer torque. The spin-orbit torque can have various contributions~\cite{Brataas.Kent.eaNM2012,PhysRevMaterials.3.011401} depending on the underlying symmetry with the simplest form, $\boldsymbol \tau_{so}=\tau_1(\hat{z}\times\boldsymbol j_c)\times\boldsymbol m +\tau_2\boldsymbol m\times[(\hat{z}\times\boldsymbol j_c)\times\boldsymbol m]$, where $\tau_1$ and $\tau_2$ describe the efficiency of the field-like and the damping-like contributions, respectively. Instabilities in the LLG equation can be revealed by first studying the spin wave modes in the absence of dissipative terms in the LLG equation, i.e., we initially put $\alpha=0$ and $\beta=0$. 

To further analyze the system, we construct the Bogoliubov-de Gennes Hamiltonian, as in \cite{Muller_2016}, but with extra terms corresponding to the current and additional terms in DMI. To this end, we describe fluctuations around the equilibrium magnetization by employing a complex field $\psi(x,y,t)$ such that $\hat{\boldsymbol m}=\hat{\bf e}_{3}\sqrt{1-2\mid \psi \mid^{2}}+\hat{e}_{+}\psi+ \hat{e}_{-}\psi^{*}$, where $\hat{e}_{\pm}=(\hat{e}_{1}\pm i \hat{e}_{2})/\sqrt{2}$ and $\hat{\bf e}_{1}^T=[-\sin\phi,\cos\phi,0]$, $\hat{\bf e}_{2}=\hat{\bf e}_{3} \times \hat{\bf e}_{1}$, $\hat{\bf e}_{3}^T=[\sin\theta(x)\cos\phi,\sin\theta(x)\sin\phi,\cos\theta(x)]$. Note that we only consider situations in which $\Tilde{D}=0$, which determines the angle $\phi$ in the parametrization of spin waves. In the general case analytical expressions become complicated, e.g., see discussion in the previous subsection. We now expand Eq.~(\ref{LLG}) up to the lowest order in the field $\psi$, arriving at the Bogoliubov-de Gennes Hamiltonian describing the mixing of the circular modes,
\begin{equation}
H_{BdG}\Psi =i\tau^{z}\partial_t \Psi,
\end{equation}
where $\Psi=( \psi,\psi^{*})^{T}$ and $\boldsymbol \tau$ stands for Pauli matrices in the Nambu space. As the system is translationally invariant along the $y$-axis, we apply the Fourier transform, arriving at the expression:  
\begin{equation}\label{BdGH0}
H^{(i)}_{BdG}=-J^{(i)}\partial^{2}_{x} +J^{(i)}q^2_{y} + H^{(i)}+2 K^{(i)}+V^{(i)}(x,q_y),
\end{equation}
where the combination $H^{(i)}+2 K^{(i)}$ defines the bulk magnon gap, and
\begin{widetext}
\begin{eqnarray}
\label{BdGVani}
    V^{(i)}(x, q_{y})=&&\mathbb{1}\bigg(-2 K^{(i)} \sin^{2}\theta + \theta^{\prime} (D^{(i)}_2 \cos{\phi}-D^{(i)}_1 \sin{\phi}-J^{(i)}\theta^{\prime}) \bigg)\\+&&\tau^{x}\bigg(K^{(i)}\sin^{2}\theta+\theta^{\prime}(D^{(i)}_2\cos{\phi}-D^{(i)}_1\sin{\phi}-J^{(i)}\frac{\theta^{\prime}}{2})\bigg)\nonumber 
        +\tau^{z}q_{y}\bigg(-2 (D^{(i)}_3\cos{\phi}+D^{(i)}_4\sin{\phi})\sin\theta+\frac{\hbar {\cal P}}{2 e}j_c\bigg),
\end{eqnarray}
\end{widetext}
where current $j_c$ is along the $y$-axis and $q_y$ is the momentum along the $y$-axis.
After accounting for boundary conditions, Eq.~\eqref{BdGH0} is solved numerically to find the magnon spectrum, e.g., as shown in Fig.~\ref{fig:modes}.

In deriving Eq.~\eqref{BdGH0}, we disregarded the spin-orbit torque. We expect qualitatively similar behavior in the presence of the spin-orbit torque where this torque can modify the bulk gap and the magnetization profile. Furthermore, the spin-orbit torque can modify the skyrmion and antiskyrmion Hall effects as discussed in the next section.

\section{Results and discussion}
We note that results in Section III hold for both regular and in-plane skyrmions and antiskyrmions as the mapping in Section IIA is also applicable to the spin-transfer torque term in the LLG equation~\cite{PhysRevB.93.064428,PhysRevB.99.060407,PhysRevB.101.054405}, while the spin-orbit torque modifications can be easily included.

\subsection{Magnon spectrum and boundary instabilities}
We first study the spin wave modes in the absence of dissipative terms in the LLG equation. Some of the eigenmodes described by Eq.~(\ref{BdGH0}) are bound to the edge and have energies within the bulk magnon gap. Such bound solutions decay into the bulk, and they can be characterized by the number of nodes in the wave functions.
In Fig.~\ref{fig:modes}, we plot the magnon spectrum of the lowest energy mode localized at the edge for different values of charge current. The presence of DMI leads to non-reciprocity in the magnon spectrum. The contribution of the current term in the Bogoliubov-de Gennes Hamiltonian can be interpreted as the Doppler shift effect on the magnon spectrum~\cite{Vlaminck.BailleulS2008} given by $j_c q_y/j_0$ in dimensionless units (see Fig.~\ref{fig:modes}). The Doppler shift can lead to the closure of the gap for the magnon modes localized at an edge~\cite{Muller_2016} or a boundary~\cite{PhysRevMaterials.2.124401}. As the bulk magnon gap is still open, this can lead to instabilities localized specifically at the boundary, and further appearance of a chiral domain wall. In micromagnetic simulations with realistic material parameters, we confirm the appearance of chiral domain walls, which eventually turn into topological defects. 
\begin{figure}
\centering
\includegraphics[width=\linewidth]{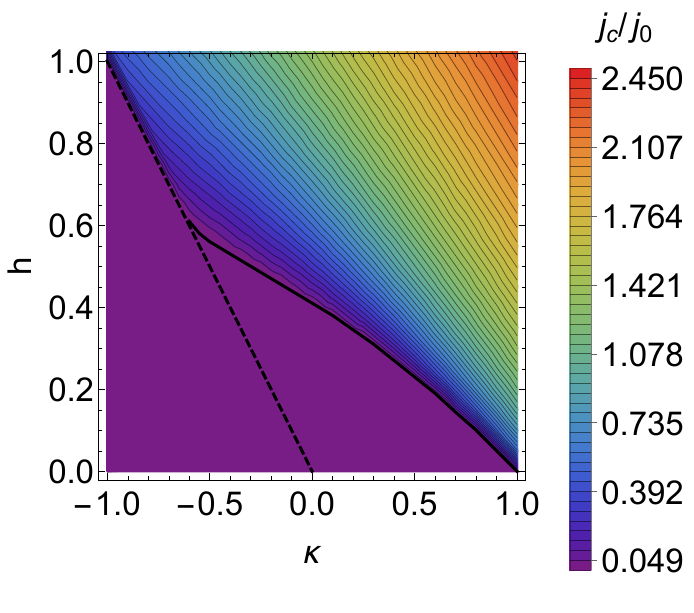}
\caption{(Color online) The diagram in ($\kappa$, $h$) space identifying instabilities associated with closure of the magnon gap at an edge. We assume $|D_4 D_1-D_3 D_2|=|{\cal D}_1|^2$ which is satisfied for DMI of the Rashba or Dresselhaus type~\cite{10.3389/fphy.2018.00098}. The dashed black line corresponds to the closure of the bulk gap. The bold black line corresponds to the closure of the gap at zero current. Away from these lines the gap is closed in the presence of current $j_c/j_0$ shown on the right with a color bar. The same diagram also describes the closure of the magnon gap at a boundary between regions with Rashba- and Dresselhaus-type DMI. This diagram applies to skyrmions and antiskyrmions of both perpendicular and in-plane configuration, depending on the choice of $D_1, D_2, D_3,$ and $D_4$.}
\label{fig:fig4}
\end{figure}

By analyzing Eq.~(\ref{BdGVani}), we can see that the edge magnon band gap will depend on the form of the kink solution $\theta$, on the uniaxial anisotropy $\kappa$, magnetic field $h$, DMI, and charge current along the boundary, $j_c$. Thus, the edge magnon band gap can be tuned via the application of external magnetic field and charge current. In the following, we will consider how a pulse of charge current and magnetic field can close the edge magnon band gap in systems with DMI and lead to generation of skyrmions or antiskyrmions.

\begin{figure}
\centering
\includegraphics[width=\linewidth]{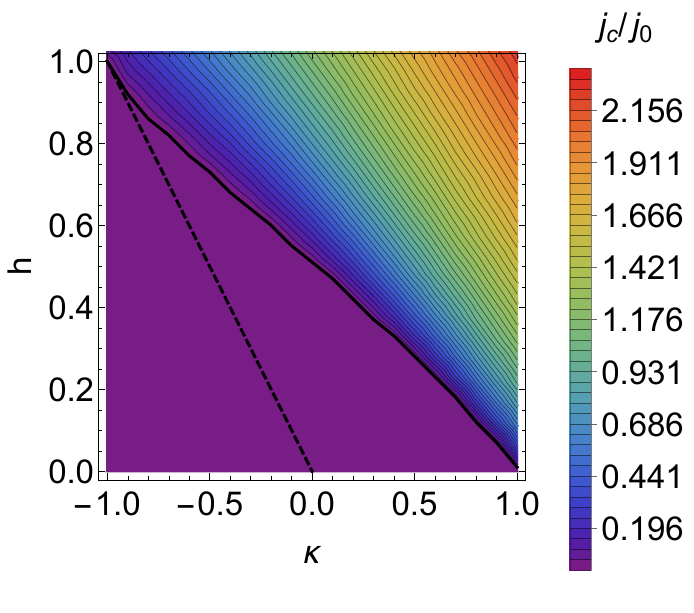}
\caption{(Color online) Same as Fig.~\ref{fig:fig4} but for parameters $|D_4 D_1-D_3 D_2|=|1.2{\cal D}_1|^2$. The same diagram also describes the closure of magnon gap at a boundary between regions with $D_1=D_4=0$, with $D_2$ changing sign across the boundary and $|D_3|=1.2|D_2|$. This diagram applies to skyrmions and antiskyrmions of both perpendicular and in-plane configuration, depending on the choice of $D_1, D_2, D_3,$ and $D_4$.}
\label{fig:fig5}
\end{figure}

\begin{figure}
\centering
\includegraphics[width=\linewidth]{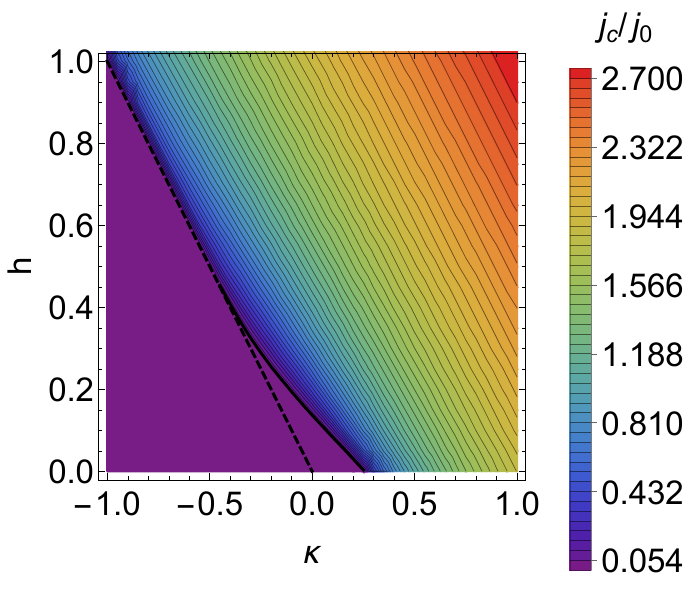}
\caption{(Color online) Same as Fig.~\ref{fig:fig4} but for a boundary between a region with vanishing DMI and a region for which $|D_4 D_1-D_3 D_2|=|{\cal D}_1|^2$. To compare with Fig.~\ref{fig:fig4}, here we use the units corresponding to $Q=\sqrt{( D_1)^2+(D_2)^2}/J$, $h=H/[J Q^2]$, and $\kappa=2K/[J Q^2]$ where only DMI changes across the boundary. This diagram applies to skyrmions and antiskyrmions of both perpendicular and in-plane configuration, depending on the choice of $D_1, D_2, D_3,$ and $D_4$.}
\label{fig:fig6}
\end{figure}

In Fig.~\ref{fig:fig4}, we explore boundary instabilities in the presence of current flowing along an edge in a system with DMI. The DMI can be, e.g., of the Rashba or Dresselhaus type, but can also be of a more general form as long as the condition $|D_4 D_1-D_3 D_2|=|{\cal D}_1|^2$ is satisfied. At zero current, we identify a large region in which the gap will only close at the boundary but not in the bulk. The color bar on the right side of the figure represents the magnitude of current needed to close the magnon band gap. We observe that the presence of charge current can expand the region of instabilities. We note that the same diagram also describes the closure of magnon gap at a boundary between regions with Rashba- and Dresselhaus-type DMI. 
In Fig.~\ref{fig:fig5}, we explore boundary instabilities for systems with anisotropic DMI.
We observe that anisotropic DMI can expand the region of instabilities, thus potentially helping in creating skyrmions and antiskyrmions at smaller currents.  Such anisotropic DMI can arise in systems with $C_{2v}$ symmetry~\cite{PhysRevB.93.064428, Hoffmann}. 
In Fig.~\ref{fig:fig6}, we consider a boundary between two regions where DMI is present for $x>0$ and DMI vanishes for $x<0$. As in Fig.~\ref{fig:fig4}, the DMI can be, e.g., of the Rashba or Dresselhaus type, but can also be of a more general form as long as the condition $|D_4 D_1-D_3 D_2|=|{\cal D}_1|^2$ is satisfied. We observe a much smaller region of boundary instabilities compared to Fig.~\ref{fig:fig4}, which can be explained by a smaller equilibrium twist of magnetization at the boundary.
According to our micromagnetic simulations, the instabilities caused by the closure of the magnon gap can lead to generation of skyrmions or antiskyrmions. We observe a qualitative agreement between the instability regions shown in the diagrams and the parameters in our micromagnetic simulations leading to generation of skyrmions or antiskyrmions. However, we also observe that the choice of dissipative parameters, $\alpha$ and $\beta$, can strongly influence our simulations, as explained in the following subsections.
\begin{figure}
\centering
\includegraphics[width=\linewidth]{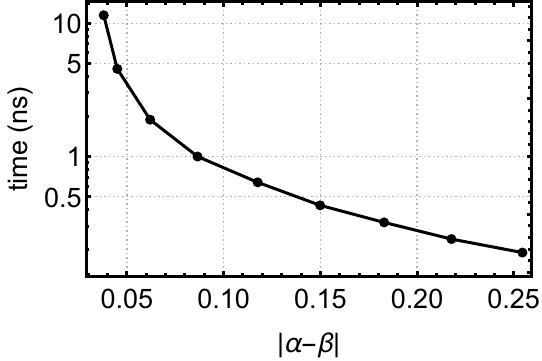}
\caption{Log plot of minimum duration of current pulse for generation of at least one topological defect from an edge in nanoseconds versus $|\alpha-\beta|$. The effective DMI strength is $D=3.0$ mJ/m\textsuperscript{2}. Other parameters in dimensionless units are $\kappa=0.2$, $h=0.63$, and $j=0.89$ (corresponding to $K=3.0 \times 10^4$ J/m\textsuperscript{3}, $H=0.326$ T, and $j_c=8.12 \times 10^{12}$ A/m\textsuperscript{2}).}
\label{fig:RashbaAlpha}
\end{figure}

\subsection{Hall effect of skyrmions and antiskyrmions} 
The application of charge current can lead to the Hall effect of skyrmions and antiskyrmions~\cite{Jonietz.Muehlbauer.eaS2010,Yu.Kanazawa.eaNC2012,Sampaio2013,Iwasaki.Mochizuki.eaNN2013,Woo.Song.eaNC2017,PhysRevB.93.064428,10.3389/fphy.2018.00098,PhysRevB.96.144412,PhysRevB.93.064428}. In the steady flow regime, such a Hall effect will depend on the dissipative parameters $\alpha$ and $\beta$, and the topological charge, as it follows from the Thiele equation. With the help of the Hall effect, the topological defects can be pushed away from the boundary by properly choosing the current direction, thus facilitating the generation of topological defects. In the steady flow regime, the dynamics of skyrmions and antiskyrmions in response to currents and potential-type forces can be described by the Thiele equation~\cite{10.3389/fphy.2018.00098}: 
\begin{align}
  s( {\cal Q} \hat z\times + \alpha \hat\eta) \boldsymbol v  =\frac{1}{4 \pi}\boldsymbol {\cal F} ,\label{force}
\end{align}
where ${\cal Q} = \frac{1}{4 \pi}\int d^2 r\,  \boldsymbol m \cdot(\partial_x \boldsymbol m \times \partial_y \boldsymbol m)$ is the topological charge,
$\boldsymbol v$ is the velocity of the topological defect, $\hat \eta$ is the damping dyadic tensor, and $\boldsymbol {\cal F}=\boldsymbol{\cal F}_{so}+\boldsymbol{\cal F}_{st}+\boldsymbol{\cal F}_{b}$ is the total force acting on the skyrmion or antiskyrmion due to the spin-orbit torque, the spin-transfer torque, and the boundary potential, respectively.
Note that for in-plane (anti)skyrmions in coordinates in Fig.~\ref{fig:IntroBimeron}, we need to perform an operation sending $x$ to $x$, $y$ to $z$, and $z$ to $-y$. Equation~\eqref{force} can describe anisotropies in response to charge currents through tensor $\hat\eta$ (e.g., due to elongation of skyrmions and antiskyrmions)~\cite{PhysRevB.93.064428}, and spin-orbit torque (e.g., for antiskyrmions)~\cite{PhysRevB.96.144412}. The latter can be described by linear relations,  
$\boldsymbol {\cal F}_{so}=4 \pi\hat{\mathcal{B}}_{so}\cdot \boldsymbol j_c$ where (in general anisotropic) tensor $\hat{\mathcal{B}}_{so}$ is proportional to $\tau_2$ (or the spin Hall angle) and is determined by the configuration of the skyrmion or antiskyrmion~\cite{PhysRevB.96.144412}. 
Similarly, for the spin-transfer torque we write $\boldsymbol {\cal F}_{st}=4 \pi\hat{\mathcal{B}}_{st}\cdot \boldsymbol j_c$ where $\hat{\mathcal{B}}_{st}=-\hbar {\cal P}/(2 e)( {\cal Q} \hat z\times + \beta \hat\eta)$. 
Equation~\eqref{force} leads to the velocity of skyrmions or antiskyrmions:
\begin{align}
v_x =\frac{1}{4\pi s} \frac{{\cal Q} {\cal F}_y +  \alpha \eta_2 {\cal F}_x}{ {\cal Q}^2 + \alpha^2 \eta_1\eta_2}, \, v_y = \frac{1}{4\pi s}   \frac{- {\cal Q} {\cal F}_x +  \alpha \eta_1 {\cal F}_y }{ {\cal Q}^2 + \alpha^2 \eta_1\eta_2},\label{eq:vel}
\end{align}
and to the Hall response described by the Hall angle:
\begin{equation}
    \theta_H = \tan^{-1}(v_y / v_x)=\tan^{-1}\left(\frac{- {\cal Q} {\cal F}_x +  \alpha \eta_1 {\cal F}_y}{{\cal Q} {\cal F}_y +  \alpha \eta_2 {\cal F}_x}\right),\label{eq:vel1}
\end{equation}
where Eqs.~\eqref{eq:vel} and \eqref{eq:vel1} are written in a reference frame in which the tensor $\hat\eta$ is diagonal with the diagonal elements $\eta_1$ and $\eta_2$. 

As follows from Eq.~\eqref{eq:vel}, the spin-orbit and spin-transfer torques can be used to facilitate the generation of topological defects by pushing defects away from the boundary for properly chosen current direction. Below, we study this effect in detail for the spin-transfer torque, while in the presence of the spin-orbit torque we observe qualitatively similar behavior. We consider a boundary at $x=0$ and assume a charge current along the boundary. The $x$-component of the velocity due to the spin-transfer torque becomes:
\begin{align}\label{velocity}
v_x=\frac{\alpha  - \beta }{s({\cal Q}^2 + \alpha^2 \eta^2)}\eta {\cal Q}j_c,
\end{align}
where we take an isotropic tensor $\eta$. From Eq.~\eqref{velocity} it is clear that given a properly chosen current direction, the skyrmion or antiskyrmion Hall effect can always facilitate generation of topological defects. Furthermore, when opposite types of topological defects are preferred for $x<0$ and $x>0$, i.e., skyrmions and antiskyrmions, the generation of topological defects on both sides of the boundary becomes possible. This can be realized for a boundary between the Rashba- and Dresselhaus-type DMI, as in this case the topological charge of preferred defects changes sign across the boundary. 

\subsection{Micromagnetic simulations} 
To generate skyrmions or antiskyrmions, it may not be sufficient to cross the phase boundaries
due to a possible formation of a metastable state under adiabatic change of parameters at a low enough temperature. To overcome this obstacle, we employ local instabilities in order to inject chiral solitons into the system through edges or boundaries. As we show below, these chiral solitons can be further broken into skyrmions or antiskyrmions by magnetic field and charge current pulses.

To confirm the importance of the skyrmion and antiskyrmion Hall effects for the topological defect generation, we carried out micromagnetic simulations in which the Hall effect is due to the spin-transfer torque. We use an amended Mumax3 code \cite{Vansteenkiste2014} to run micromagnetic simulations to demonstrate skyrmion and antiskyrmion generation on edges or boundaries. We use Object Oriented Micromagnetic Framework (OOMMF) to produce images of our micromagnetic results \cite{oommf}. In all simulations, we use $J/2=15$ pJ/m. We note that the values we use for DMI, exchange stiffness, uniaxial anisotropy, and saturation magnetization are in line with the values of a Co/Pt interface \cite{PhysRevB.96.060410,Sampaio2013}.
We apply periodic boundary conditions in the $y$-direction and open boundary conditions in the $x$-direction for our simulations, and we insert a notch at the edge of the magnetic region or at the boundary in order to break translational symmetry in the $y$-direction. 
We note that the notch is not necessary if we use open boundary conditions in both the $x$- and $y$-direction, as in that case the translational symmetry in the $y$-direction is broken by the finite size of the system. We also note that to check the correctness of boundary conditions in the amended Mumax3 code, we have compared the profile of the magnetization at the edge of a region with DMI and at a boundary between regions of differing DMI from numerics and the analytical solutions, and we have found perfect agreement.

\begin{figure}
\centering
\includegraphics[width=\linewidth]{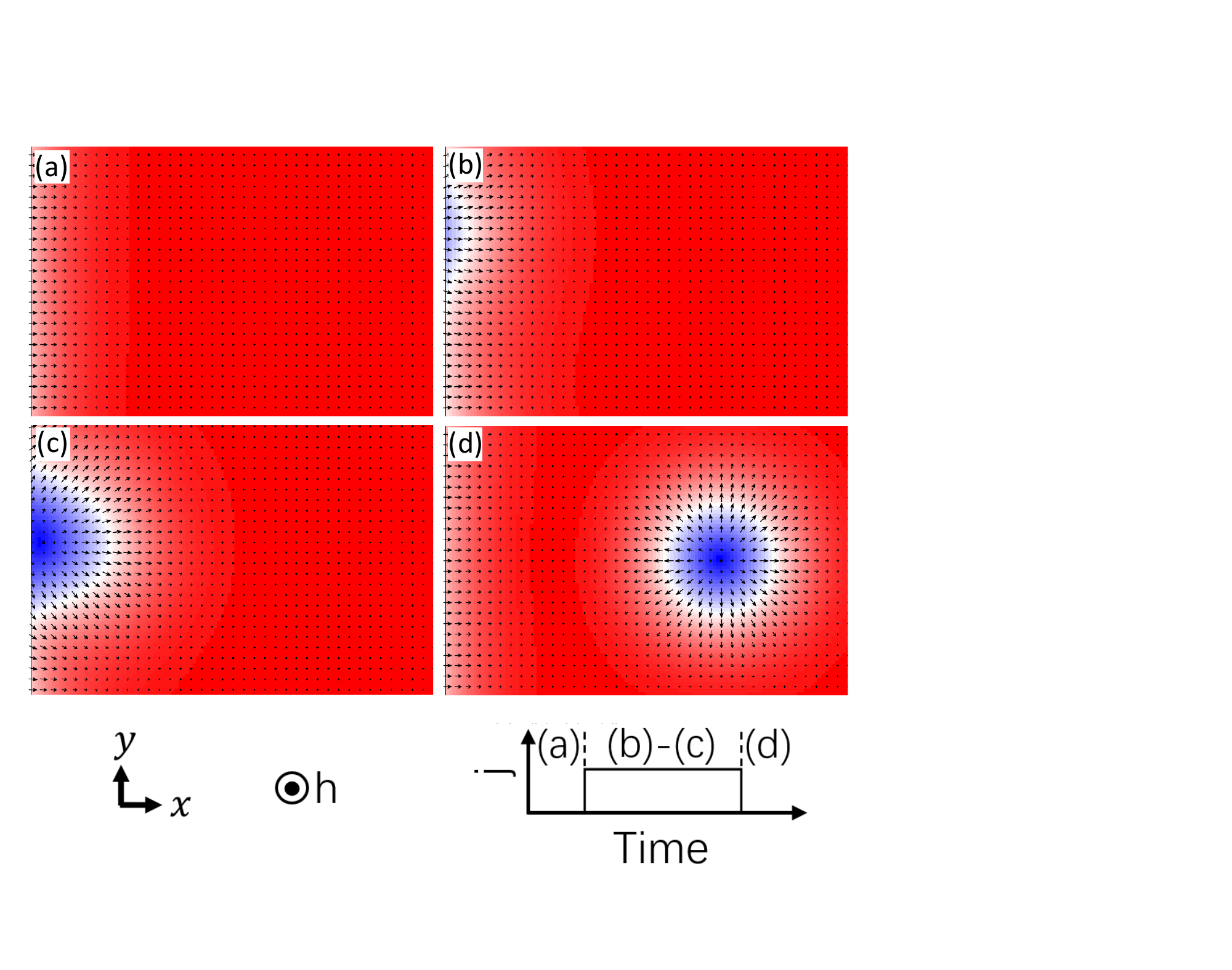}
\caption{(Color online) Skyrmion generation at the edge of a region with Rashba-type DMI. Red shading denotes magnetization in the $+z$-direction and blue shading denotes magnetization in the $-z$-direction. (a) The magnetic texture is relaxed in the presence of magnetic field $h=0.7$. (b)-(c) A current is applied and magnetic field is lowered to $h=0.4$ for a time of $\Delta t=4.25$~ns. (d) After the current is turned off and the magnetic field is returned to its original value, the skyrmion is stable. The following parameters have been used: $\kappa=0.2$, $j_c/j_0=0.216$ (corresponding to $K=3.0 \times 10^4$~J/m\textsuperscript{3} and $j_c=1.97 \times 10^{12}$ A/m\textsuperscript{2}).}
\label{fig:SkCreation}
\end{figure}

\begin{figure}
\centering
\includegraphics[width=\linewidth]{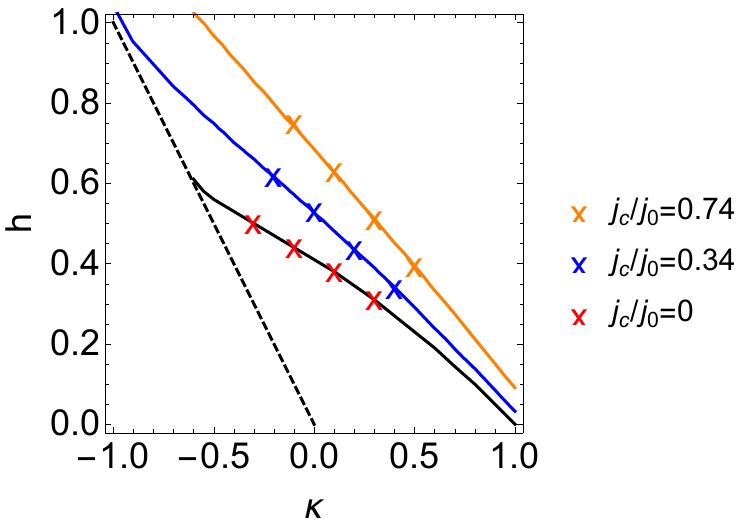}
\caption{(Color online) Results of micromagnetic simulations of skyrmion generation at the edge of a region with Rashba-type DMI (cf. Fig.~\ref{fig:fig4}). The protocol is similar to that of Fig.~\ref{fig:SkCreation}: the magnetic texture is relaxed in the presence of a magnetic field, the magnetic field is lowered and current is applied (to the values shown by the crosses), and after current is turned off, the magnetic field is returned to its original value. We use $D=3.0$ mJ/m\textsuperscript{2} (corresponding to $d_{3}=-d_{2}=1$ in dimensionless units), $M_s=5.8 \times 10^5$ Am\textsuperscript{-1}, $\alpha=0.03$, and $\beta=0.09$. We use a notch with a radius of $5$ nm on the left side of the magnetic region. The solid lines show current values from Fig.~\ref{fig:fig4}, with black for $j_c/j_0$=0, blue for $j_c/j_0$=0.34, and orange for $j_c/j_0$=0.74.}
\label{fig:fig8A}
\end{figure}

\begin{figure}
\centering
\includegraphics[width=\linewidth]{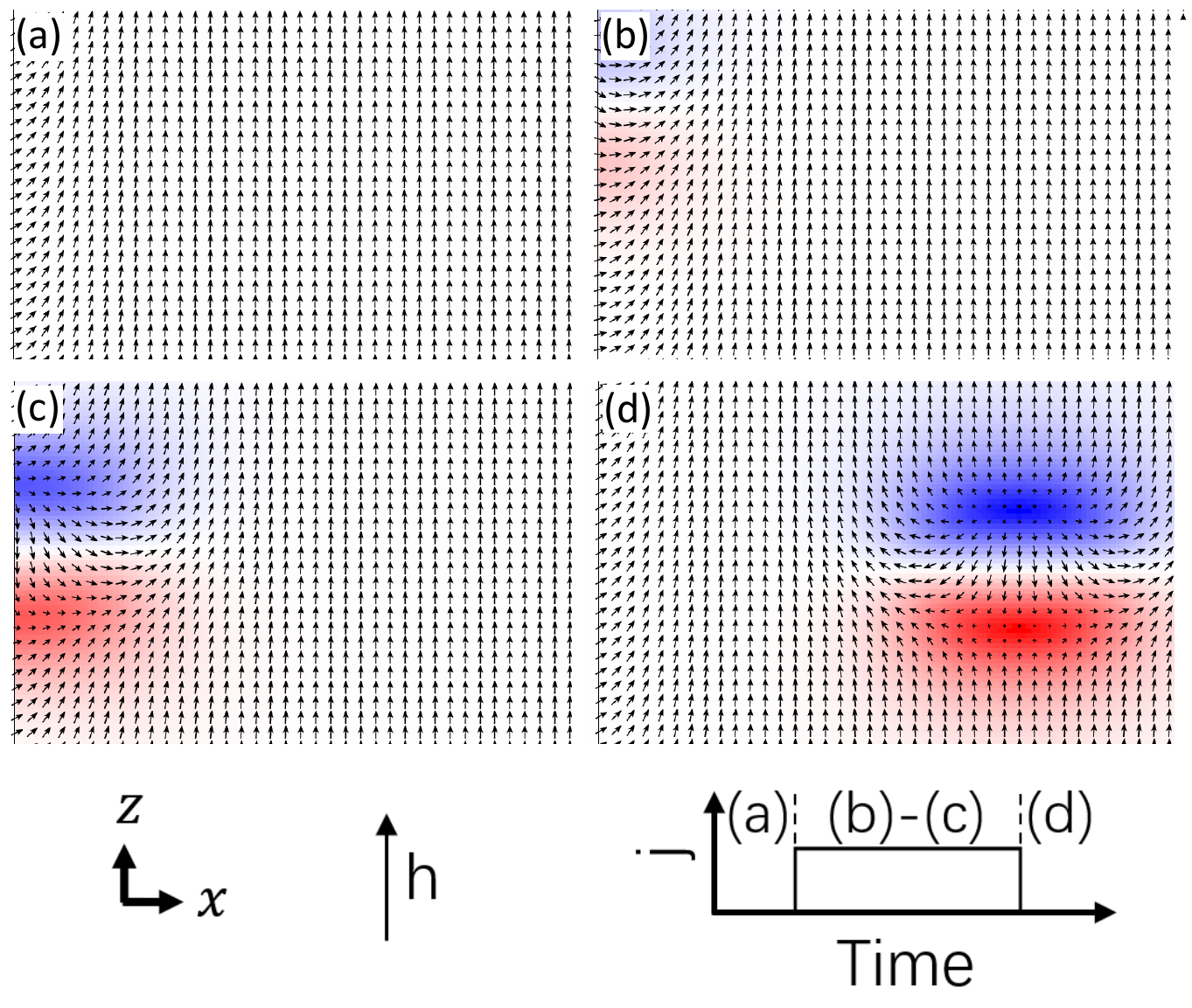}
\caption{(Color online) In-plane skyrmion generation at the edge of a region with parameters $D_2=-3.0$~mJ/m\textsuperscript{2} and $D_3=-1.2 D_2$. Red shading denotes magnetization in the $-y$-direction and blue shading denotes magnetization in the $+y$-direction. (a) The magnetic texture is relaxed in the presence of magnetic field $h=0.95$. (b)-(c) A current is applied and magnetic field is lowered to $h=0.58$ for a time of $\Delta t=6.45$~ns. (d) After the current is turned off and the magnetic field is returned to its initial value, the in-plane skyrmion is stable. The following parameters have been used: $\kappa=0$ and $j_c/j_0=0.196$ (corresponding to $j_c=1.79 \times 10^{12}$~A/m\textsuperscript{2}).}
\label{fig:BIMERONGRID}
\end{figure}

In the results shown below, we turn the demagnetizing field off, but we note that demagnetizing effects can be effectively included in magnetic anisotropy parameter $K$ as an additional in-plane shape anisotropy. By performing micromagnetic simulations with the demagnetizing field turned on, we verified that we can qualitatively reproduce our results by changing the parameter $K$.

By choosing the magnitude of a current pulse according to Figs.~\ref{fig:fig4}--\ref{fig:fig6} in our micromagnetic simulations, we find that the skyrmion or antiskyrmion Hall effect can facilitate the generation of topological defects by charge currents. In Fig.~\ref{fig:RashbaAlpha}, we show the minimal duration of a current pulse necessary for generation of a topological defect while keeping the magnetic field constant.
We see that generation times decrease as the difference between $\alpha$ and $\beta$ increases, as this increases the skyrmion or antiskyrmion Hall angle, which helps push the topological defects away from the edge or boundary. We obtain qualitatively similar results for different values of $\alpha$ and $j_c$, but in general as $\alpha$ increases, all generation times increase, and as $j_c$ increases, all generation times decrease.

In Fig.~\ref{fig:SkCreation}, we simulate a 256~nm $\times$ 256~nm region with Rashba-type DMI. We use $D=3.0$ mJ/m\textsuperscript{2} (corresponding to $d_{3}=-d_{2}=1$ in dimensionless units), $M_s=5.8 \times 10^5$ Am\textsuperscript{-1}, $\alpha=0.03$, and $\beta=0.09$. We use a notch with a radius of $5$ nm on the left side of the magnetic region. An external magnetic field is applied in the positive $z$-direction. For a time period of $\Delta t=4.25$ ns, a charge current pulse is applied in the positive $y$-direction and the magnetic field is lowered. The current causes instabilities at the edge, which turn into a chiral domain pushed by the current in the positive $x$-direction. The charge current is then turned off and the magnetic field returned to its initial value. In Fig.~\ref{fig:SkCreation}, we show results for such a protocol. We note that, as $j_c$ is applied in the positive $y$-direction, and $\beta>\alpha$, we find that $v_x>0$ due to the Hall effect, according to Eq.~\eqref{velocity}, and so we see generation of skyrmion from the left edge of the magnetic region. If we use $\beta<\alpha$, generation occurs only when $j_c$ is applied in the negative $y$-direction. We have also sampled various points in the diagram in Fig.~\ref{fig:fig4} to confirm the agreement with our analysis of instabilities. These results are shown in Fig.~\ref{fig:fig8A}

\begin{figure}
\centering
\includegraphics[width=\linewidth]{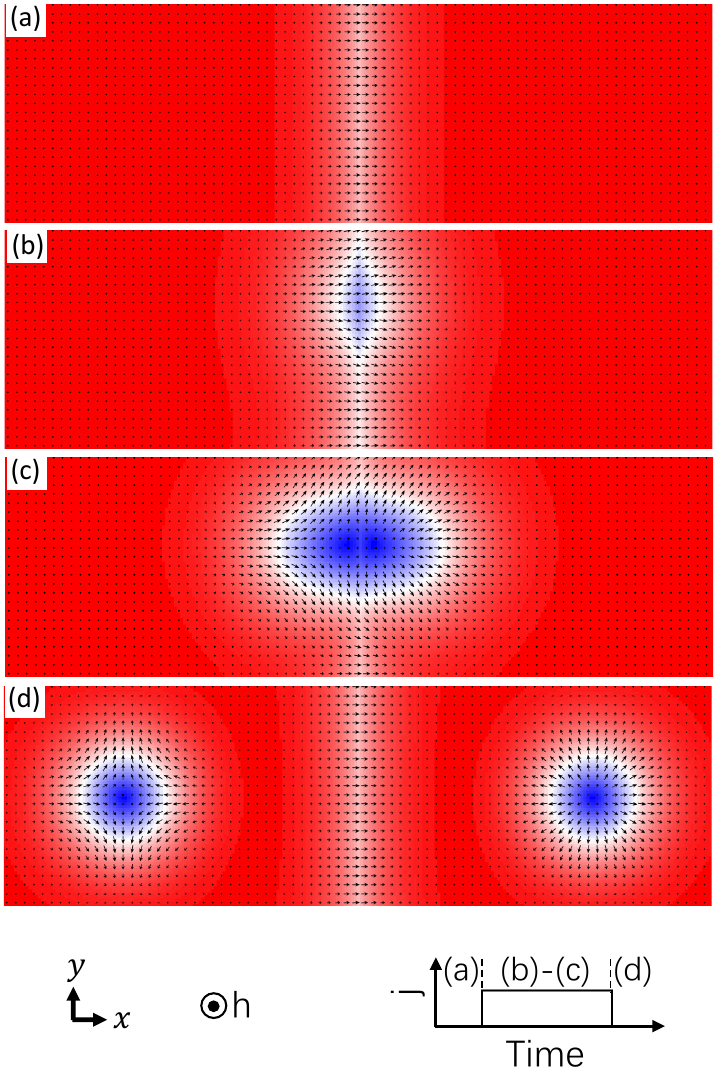}
\caption{(Color online) Antiskyrmion and skyrmion generation at a boundary between the Dresselhaus-type DMI on the left and the Rashba-type DMI on the right. Red shading denotes magnetization in the $+z$-direction and blue shading denotes magnetization in the $-z$-direction. (a) The magnetic texture is relaxed in the presence of magnetic field $h=0.83$. (b)-(c) A current is applied and magnetic field is lowered to $h=0.48$ for a time of $\Delta t=5.25$ ns. (d) After the current is turned off and the magnetic field is returned to its initial value, the antiskyrmion and skyrmion are stable. The following parameters have been used: $\kappa=0$ and $j_c/j_0=0.21$ (corresponding to $j_c=1.92~\times 10^{12}$ A/m\textsuperscript{2}).}
\label{fig:INTGRID}
\end{figure}

In Fig.~\ref{fig:BIMERONGRID}, we simulate the generation of in-plane skyrmions from an edge. We simulate a 256 nm $\times$ 256 nm region. We use $D_2=-3.0$ mJ/m\textsuperscript{2}, $D_3=-1.2 D_2$, $M_s=5.8 \times 10^5$ Am\textsuperscript{-1}, $\alpha=0.03$, and $\beta=0.09$. The anisotropic DMI~\cite{Camosi.Rohart.ea:PRB2017} should lead to elongation of skyrmions along the $x$-axis~\cite{PhysRevB.93.064428}, which is confirmed in Fig.~\ref{fig:BIMERONGRID}. We simulate a notch with a radius of 5 nm on the left side of the magnetic region. An external magnetic field is applied in the positive $z$-direction. For a time period of $\Delta t=6.45$ ns, a charge current pulse is applied in the positive $z$-direction and the magnetic field is lowered. Note that mechanisms of generation of skyrmions in Figs.~\ref{fig:SkCreation} and \ref{fig:BIMERONGRID} are identical, which can be best seen by performing a rotation around the $x$-axis described in Fig.~\ref{fig:IntroBimeron}, as this rotation also preserves the form of the spin-transfer torque term. For the spin-orbit torque term, the skyrmion or antiskyrmion Hall effect is modified according to Eq.~\eqref{eq:vel} in response to a rotation around the $x$-axis.
We have sampled various points in the diagram in Fig.~\ref{fig:fig5} to confirm the agreement with our analysis of instabilities.

In Fig.~\ref{fig:INTGRID}, we simulate the generation of skyrmions and antiskyrmions at a boundary with Dresselhaus-type DMI on the left and Rashba-type DMI on the right. We simulate a 512 nm $\times$ 256 nm region. We use $D=3.0$~mJ/m\textsuperscript{2} (corresponding to $d^{(1)}_3=d^{(1)}_2=1$ and $d^{(2)}_3=-d^{(2)}_2=1$), $M_s=5.8 \times 10^5$ Am\textsuperscript{-1}, $\alpha=0.03$, and $\beta=0.09$. We simulate a notch with a radius of 5 nm in the center of the magnetic region. An external magnetic field is applied in the positive $z$-direction. For a time period of $\Delta t=5.25$ ns, a charge current pulse is applied in the positive $y$-direction and the magnetic field is lowered. In Fig.~\ref{fig:INTGRID}, we show results for such a protocol. We note that, as $j_c$ is applied in the positive $y$-direction, and $\beta>\alpha$, we find that skyrmions and antiskyrmions are pushed away from the boundary due to their opposite topological charge, in agreement with Eq.~\eqref{velocity}, and so we see generation of topological defects at the boundary. If we use $\beta<\alpha$, generation occurs only when $j_c$ is applied in the negative $y$-direction. We have sampled various points in the diagram in Fig.~\ref{fig:fig4} to confirm the agreement with our analysis of instabilities.

\section{Summary} We have formulated a theory of regular and in-plane skyrmion and antiskyrmion generation at edges of magnetic films and at boundaries between regions with different DMI. The process of generation is triggered by local instabilities at edges or boundaries due to lowering of magnetic field or application of charge current pulse. To identify the appearance of instabilities, we have studied the magnon modes localized at edges
or boundaries. In our micromagnetic simulations, the magnon gap for such modes closes while the bulk magnon gap is still finite. As a result, the generation only happens at edges or boundaries. By studying a charge current flowing along the edge or boundary, we have concluded that the presence of the skyrmion or antiskyrmion Hall effect can facilitate the generation of topological defects. Depending on the direction of charge current, topological defects are pushed either away from or towards the edge or boundary, which either facilitates or suppresses the generation of skyrmions or antiskyrmions. 

In our micromagnetic simulations, we have also studied the effects of dipolar interactions where we have accounted for both the surface and the volume magnetic charges. We have found that our approach of skyrmion or antiskyrmion generation also works in the presence of dipolar interactions. We have confirmed that in a quasi-two-dimensional
geometry for antiskyrmions, dipolar interactions originating in the magnetic volume charges increase the size and can provide additional stability~\cite{PhysRevB.97.134404}, and for in-plane skyrmions and antiskyrmions, dipolar interactions originating in the magnetic volume charges diminish the average size and lead to elongation~\cite{PhysRevB.101.054405}. 

Realizations of antiskyrmions and in-plane (anti)skyrmions will require careful material engineering as the former can be realized in systems with $D_{2d}$ or $C_{2v}$ symmetry and the latter in systems with only mirror symmetry, $M_x$. We note that regular skyrmions can be hosted in Ir/Fe/Co/Pt multilayers \cite{Soumyanarayanan2017}, and regular antiskyrmions can be hosted in Heusler compounds of $D_{2d}$ symmetry \cite{Nayak2017}. For in-plane skyrmions, proposed material candidates include FeLa$_3$S$_6$ and Rb$_6$Fe$_2$O$_5$ \cite{PhysRevB.101.054405}. Systems based on magnetic heterostructures of different layered materials can in principle be engineered to realize various topological defects discussed in our work.

\section{Acknowledgments}
We gratefully acknowledge useful discussions with Bo Li and Kirill Belashchenko. This work was supported by the U.S. Department of Energy, Office of Science, Basic Energy Sciences, under Award No. DE-SC0021019. Part of this work was also completed utilizing the Holland Computing Center of the University of Nebraska, which receives support from the Nebraska Research Initiative.

\bibliographystyle{apsrev4-1}
\bibliography{lit}

\end{document}